\documentclass[showpacs,article]{revtex4}%
\usepackage{amsmath}
\usepackage{graphicx}
\usepackage{epsfig}
\usepackage{dcolumn}
\usepackage{bm}
\usepackage{dcolumn}
\usepackage{amsfonts}
\usepackage{amssymb}%
\providecommand{\U}[1]{\protect\rule{.1in}{.1in}}
\begin{document}
\title{Newton-like equations for a radiating particle }
\author{A. Cabo Montes de Oca$^{\ast}$ and N. G. Cabo Bizet$^{\ast\ast}$ \bigskip}
\affiliation{$^{\ast}$ Departamento de F\'{\i}sica Te\'{o}rica, Instituto de
Cibern\'{e}tica Matem\'{a}tica y F\'{\i}sica (ICIMAF), Calle E, No. 309, entre
13 y 15, Vedado, La Habana, Cuba}
\affiliation{$^{\ast\ast}$ Centro de Aplicaciones Tecnologicas y Desarrollo Nuclear
(CEADEN), Calle 30, esq a 5ta, Miramar, La Habana, Cuba \bigskip}

\begin{abstract}
\noindent Second order Newton equations of motion for a radiating particle are
presented. It is argued that the trajectories obeying them also satisfy the
Abraham-Lorentz-Dirac (ALD) equations for general 3D motions in the
non-relativistic and relativistic limits. The case of forces only depending of
the proper time is here considered. For these properties to hold, it is
sufficient that the external force to be infinitely smooth and that a series
formed with its time derivatives converges. This series define in a special
local way the effective forces entering the Newton equations. When the
external force vanishes in an open vicinity of a given time, the effective one
also becomes null. Thus, the proper solutions of the effective equations can
not show runaway or pre-acceleration effects. The Newton equations are
numerically solved for a pulsed force given by an analytic function along the
proper time axis. The simultaneous satisfaction of the ALD equations is
numerically checked. Further, a set of modified ALD equations for almost
everywhere infinitely smooth forces, but including step like discontinuities
in some points is also presented for the case of collinear motions. The form
of the equations supports the statement argued in a previous work, about that
the causal Lienard-Wiechert field solution surrounding a radiating particle,
implies that the effective force on the particle should instantaneously vanish
when the external force is retired. The modified ALD equations proposed in the
previous work are here derived in a generalized way including the same effect
also when a the force is instantly connected.

\bigskip

\noindent A. Cabo E-Mail: cabo@icimaf.cu \newline\noindent\ N. G .Cabo Bizet
\ E-Mail: nana@ceaden.edu.cu

\end{abstract}

\pacs{03.50.De, 02.30.Ks, 41.60.-m, 24.10.Cn, 03.65.Ta, 87.15.A-, 98.70.Rz}
\maketitle


\section{Introduction}

The long-standing search for a consistent formulation of the
Abraham-Lorentz-Dirac equation (ALD) for a radiating particle remains being a
field of intense research activity \cite{Lorentz,abraham,poincare,
schot,Dirac, Teitelboim,Landau,Rohrlich,Yighjian,Ford,Spohn,Raju}. The
complete understanding of the formulation and solution of the equations had
presented hard theoretical difficulties. The existence of solutions with
values growing without bound (runaway behavior) or pre-accelerated motions in
advance to the applied forces, are two of the most debated issues associated
to the ALD equations \cite{Yighjian,Spohn}. Important advances in the
understanding of these properties had been by now obtained in Refs.
\cite{Yighjian,Ford}. In particular, in Ref. \cite{Yighjian}, it was argued
that the ALD equations can't be derived in a completely exact way from the
integral equations satisfied by the coupled motion of a particle and its
accompanying electromagnetic field for non analytically depending forces.
\ Moreover, the author derived a correction to the ALD equations which should
be used in order to consider suddenly changing forces. The change corresponded
to multiplying the reaction force of the field on the particle by a
proper-time dependent function $\ \eta(\tau)$ which vanish at the proper time
value at which the forces is started to be applied, and attains a unit value
in a small time interval of the order of the time required by light to travel
the radius of the considered structured body studied in Ref. \cite{Yighjian}.
This change was argued to allow eliminating both: the runaway and the
pre-accelerated types of solutions. Thus, in general terms, it was concluded
that the ALD equations are not exact consequences of the integral equations of
motions for the coupled system describing the structured particle investigated
in Ref. \cite{Yighjian}. In addition, in Refs \cite{Ford}, the authors
presented a derivation of a simpler second order in the time derivative of the
position, for a radiating particle, assuming that it shows an internal
structure. The equation derived was argued to be exact for the structured
particle, which could justify its incomplete equivalence with the ALD equation
\cite{Ford}.

In the present work, at variance with the studies in Refs.
\cite{Yighjian,Ford} we will not consider that the radiating particles has a
structure. \ Our work, is rather devoted to identify and study some properties
of a curious second order Newton type of equation motion for the radiating
particle, that when satisfied, by assuming that the force is a infinitely
smooth function, directly obey the ALD equations. This occurs in a proper time
region of convergence of a particular series conformed with the infinite
sequence of the time derivatives of the force at a given instant. The
effective force at this instant results to be defined by the local values of
the series at that moment, which implies that the\ effective force vanishes
when the external one is null within an open neighborhood, which eliminates
the possibility of pre-acceleration of runway solutions when the external
force vanishes. \ \ To motivate the second order equations, firstly we present
their equivalence with the ALD ones in the non-relativistic version of them
for a general 3D motion. Next, the equations are generalized to the
relativistic case by a simple construction. The work also presents the
solution of the equations for a simple example of a pulse like force for which
the above mentioned particular series of their proper time derivatives
converge in the whole time axis. The effective force is evaluated, and the
position, velocity, acceleration and its derivative, are determined from the
numerical solution of the equations. The results does not show runaway or
pre-accelerated behaviors, since as noted before, the effective force only
depends on the local behavior around a proper time instant. The case of forces
being infinitely smooth, but eventually showing a numerable set of step like
discontinuities is also analyzed for the case in which the motion is
collinear. It becomes possible to derive modified ALD equation of motion,
after assuming that, as in the example of the analytic pulse, the
discontinuous force can be expressed as the limit of a sequence of infinitely
smooth forces, which define a corresponding sequence of effective forces also
showing the same kind of discontinuity as the external one in the limit. The
modified ALD equations give a generalization of the ones derived in reference
\cite{cabojorge} and fully support the central idea proposed in that work:
that the validity of the Lienard-Wiechert solutions for the electromagnetic
field closely around the particle, \ just after the external forces is
retired, implies that its acceleration should also suddenly disappears. This
is a natural consequence of the fact that the electromagnetic field in a
sufficiently close small vicinity of the particle is given by a Lorentz
boosted Coulomb field, which is known to not produce any force on the central
particle. Using these modified equations we also investigated their solutions
for an external force in the form of a rigorously squared pulse which exactly
vanishes outside a given time interval. An important point to stress is that
the Delta function like terms within the smooth solution for the squared pulse
obtained, follow as "normal" contributions to the usual "radiation" force
terms. Thus, their presence is consistent with the well understood notion
about that the ALD equation well describe the radiation properties of a moving
particle. However, in addition it also follows that such Delta function terms
exactly cancel with other ones associated to time derivatives of step like
changes in the acceleration occurring at the discontinuities. This means that,
effectively, the discontinuities, do not contribute with finite terms to the
radiation, contrary to what could be expected due to the appearance of the
Delta functions in the modified equations.

\bigskip

The expositions proceed as follows. In section 2, the Newton like equation
which solutions also satisfy the non-relativistic ALD equations for a general
3D dynamics is presented. Section 3, generalizes the discussion by
constructing a relativist Newton like equation which also satisfies the
relativistic ALD one. \ Next, Section 4 presents concrete conditions which
determine that given an infinitely smooth external force, the effective force
determined by it, becomes also well defined. Section 5 continues by presenting
the numerical solution of the Newton second order equations for the case of a
force being similar to a squared pulse, but being defined by an analytical
function along all the time axis. The following Section 6 is devoted to derive
the modified ALD equations for forces defined almost everywhere by infinitely
smooth functions, but which show step like discontinuities at a denumerable
set of instants along the time axis. In Section 7, the modified ALD equations
are solved for the rigorous squared pulse. This force pertains to the class of
almost everywhere $C^{\infty}$ functions, showing step like discontinuities.
The solution is compared with the one associated to the previously studied
analytical pulsed force. They show a very close appearance, indicating the
presence of Dirac delta functions concentrated at the discontinuity points of
the external force in the modified ALD equations. Finally, in the summary the
results are shortly reviewed and possible extensions of the work commented.
\bigskip

\section{The case of the non-relativistic ALD equations}

Let us consider a general but nonrelativistic motion of a particle $P$ along a
space-time trajectory $C$ defined by a curve $x$($\tau)=(x^{1}(t),\,x^{2}%
(t),\,x^{3}(t))$ and parameterized by the time $t.$ \ For this motion, the
non-relativistic ALD equations take the standard form
\begin{align}
m\text{ }a^{i}(t)-f^{i}(t)\text{\ }  &  \text{=}\,\kappa\text{ }\frac
{da^{i}(t)}{dt}\text{ },\label{ALD}\\
v^{i}(t)\text{ }  &  \text{=\ }\overset{\cdot}{x}^{i}(t)\text{ =}\frac
{dx^{i}(t)}{dt}\text{\ ,}\\
a^{i}(t)\text{ }  &  \text{=\ }\overset{\cdot}{v}^{i}(t)\text{ =}\frac
{dv^{i}(t)}{dt}\text{\ ,} \label{secondOrder}%
\end{align}
where the index $i$ has the three values $i=1,2,3$. \ In what follows, without
attempting to repeat the trial and error process which led to the mentioned
solution, we will directly write its expression for afterwards arguing that it
solves the above equations (\ref{ALD}).

The second order Newton-like equations have the specific form
\begin{equation}
a^{i}(t)=\frac{1}{m}\sum_{n=0}^{\infty}\frac{d^{n}}{dt^{n}}f^{i}%
(t)(\frac{\kappa}{m})^{n}. \label{newton}%
\end{equation}

Let us assume that the forces are infinitely smooth, that is, pertaining to
$C^{\infty}$ and that the series at the right hand side (r.h.s) converges in
certain interval of times. Then, for the time derivative of the acceleration
we have
\begin{align}
\overset{\cdot}{a}^{i}(t) &  =\frac{1}{m}\sum_{n=0}\frac{d^{n+1}}{dt^{n+1}%
}f^{i}(t)(\frac{\kappa}{m})^{n}\nonumber\\
&  =\frac{1}{\kappa}\sum_{n=0}\frac{d^{n+1}}{dt^{n+1}}f^{i}(t)(\frac{\kappa
}{m})^{n+1}\nonumber\\
&  =\frac{1}{\kappa}(\sum_{n=0}\frac{d^{n}}{dt^{n}}f^{i}(t)(\frac{\kappa}%
{m})^{n}-f^{i}(t))\nonumber\\
&  =\frac{1}{\kappa}(m\text{ }a^{i}(t)-f^{i}(t)),\label{property}%
\end{align}
where, as before, in the following a point over a quantity will mean a
derivative over the defined temporal argument of it. Therefore, assumed that
the series is well defined, the trajectory $x^{i}(t)$ \ solving equation
(\ref{newton}), also satisfies the non-relativistic ALD\ equations%
\begin{equation}
m\text{ }a^{i}(t)-f^{i}(t)=\kappa\,\overset{\cdot}{a}(t).
\end{equation}

Now, a question appears about the existence of proper and helpful definitions
of the effective force at the r.h.s of equation (\ref{newton}). \ Let us
differ the discussion of this point up to the next sections. There, we will
derive a condition to be satisfied by the external force for the effective one
to be well defined. In addition we will construct an explicit example in which
the force is infinitely smooth at all the times, allowing to calculate the
effective one. In coming section we will generalize the discussion by
determining a second order covariant equation which solution should also
satisfy the relativistic ALD equations.

\section{The case of the relativistic ALD equations}

Let us consider a force in the instant rest frame of the particle,written in
the way
\begin{equation}
f_{e}^{\text{\ }\mu}(\tau)=(0\text{ },\text{ \ }f_{e}^{i}(\tau)),
\end{equation}
where the spatial components in the rest frame $f_{n}^{i}(\tau)$ are functions
of the proper time given in the form suggested by the discussion in the
previous section
\begin{equation}
f_{e}^{\text{\ }i}(\tau)=\frac{1}{m}\sum_{m=0}^{n}\frac{d^{m}}{d\tau^{m}}%
f^{i}(\tau)(\frac{\kappa}{m})^{m},
\end{equation}
and $f^{\text{\ }i}(\tau)$ are the components of the external forces exerted
on the particle in the rest frame. \ Note that in the rest frame the zeroth
component of the external force is always equal to zero, \ then the time
derivatives of arbitrary order of these components will be automatically zero
also. \ 

Let us consider an inertial observer's inertial frame $O$ and determine \ a
\ particular Lorentz transformation that at any value of the proper time of
the particle $\tau$ \ links the coordinates of the observer's and the proper
one. \ \ For this purpose let us imagine the trajectory of the particle
$y^{\mu}(\tau)$ \ as a function of $\tau$, as seen by the observer. Then,
assuming that the particle starts moving at the origin of the observer's from
at $\tau=0,$ \ let us divide the whole proper time interval of its movement in
$N$ equal intervals of size $\epsilon=\frac{{\Large \tau}}{N}.$ \ \ Let us now
construct a Poincare inhomogeneous transformation \ which links the
instantaneous rest frame and the observer's one. \ For this purpose let us
write the general expression for a general Lorentz boost (a Lorentz
transformation without rotation)%

\begin{align}
B_{\text{ \ }\nu}^{\mu}  &  \equiv\left(
\begin{tabular}
[c]{ll}%
$\gamma$ & $\gamma$ $v^{j}$\\
$\gamma$ $v^{i}$ & $\delta_{j}^{i}+(\gamma-1)\frac{v^{i}\text{ }v^{j}}{v^{2}}$%
\end{tabular}
\ \ \ \ \right)  ,\label{boost}\\
\gamma &  =\frac{1}{\sqrt{1-v^{2}}},\text{ \ \ \ \ }v^{2}=v^{j}v^{j}.
\end{align}

Considering that the transformation is associated to an infinitesimal velocity
$dv^{i},$ the expression reduces to%

\begin{equation}
B_{\text{ }\nu}^{\mu}\equiv\left(
\begin{tabular}
[c]{ll}%
$1$ & $dv^{j}$\\
$dv^{i}$ & $\delta_{j}^{i}$%
\end{tabular}
\ \ \right)  .
\end{equation}

But defining the four-velocity of the starting frame at the proper time $\tau
$, and the corresponding increment in the differential time interval $d\tau$
as defined \ by%

\[
u^{\mu}=\binom{1}{\overrightarrow{0}},\text{ \ \ }du^{\mu}=\binom
{0}{d\overrightarrow{v}},
\]
the infinitesimal boost in the rest frame can be written in the form
\begin{equation}
B_{\text{ \ }\nu}^{\mu}(u,du)\equiv\delta_{\text{ }\nu}^{\mu}-u^{\mu}du_{\nu
}+du^{\mu}u_{\nu}.
\end{equation}

Then, after performing a Lorentz or Poincare transformation to an arbitrary
reference frame, \ the infinitesimal transformations between two successive
rest frames separated by a small proper time interval $d\tau$, that will need
to do the observer (which see the system in movement) is defined by the same
covariant formula but \ in terms of the four velocity and its increment \ in
the form \ %

\begin{align}
u^{\mu}  &  \equiv\frac{1}{\sqrt{1-v^{2}}}(1,\text{ }\overrightarrow{v}%
(\tau))=(\gamma,\text{ }\gamma\overrightarrow{v}(\tau)),\text{ \ }\\
du^{\mu}  &  \equiv(\frac{d}{d\tau}\gamma,\frac{d}{d\tau}(\text{ }%
\gamma\overrightarrow{v}(\tau)))d\tau,\\
\text{\ }\overrightarrow{v}(\tau)  &  \equiv(v^{1}(\tau),\text{ }v^{2}%
(\tau),\text{ }v^{3}(\tau)),\nonumber\\
&  =-(v_{1}(\tau),\text{ }v_{2}(\tau),\text{ }v_{3}(\tau)).
\end{align}
The metric tensor will be assumed in the convention%
\begin{equation}
g_{\text{ \ }\nu}^{\mu}\equiv\left(
\begin{tabular}
[c]{llll}%
$1$ & $0$ & $0$ & $0$\\
$0$ & $-1$ & $0$ & $0$\\
$0$ & $0$ & $-1$ & $0$\\
$0$ & $0$ & $0$ & $-1$%
\end{tabular}
\ \right)  ,
\end{equation}
and the natural system of units is being employed. \ In the rest frame, the
four-velocity takes the reduced form
\begin{equation}
u^{\mu}(\tau)=(1,0,0,0),\text{ \ \ }u^{\mu}(\tau)u_{\mu}(\tau)=1.
\end{equation}
\ 

Having the expression for the Lorentz boosts which transform (in the
observer's frame) between \ two contiguous proper references frames
(associated to a small difference of proper times $d\tau$) we can combine a
large set of infinitesimal successive Poincare transformations in the form%

\[
U(\Lambda,y(\tau))=\lim_{N\rightarrow\infty}\left(  \prod_{n=1}^{N}%
U(B(u(n\epsilon),du(n\epsilon)),dy(n\epsilon))\right)
\]
to construct a finite Poincare transformation of the form
\begin{align*}
x^{\prime\mu}  &  =\lim_{N\rightarrow\infty}\left(  \prod_{n=1}^{N}%
B(u(\frac{n}{N}\tau),du(\frac{n}{N}\tau))\right)  _{\nu}^{\mu}x^{\nu}+y^{\mu
}(\tau)\\
&  =\Lambda_{\nu}^{\mu}(\tau)x^{\nu}+y^{\mu}(\tau)\\
y^{\mu}(\tau)  &  =dy^{\nu}(\epsilon)+\sum_{m=2}^{N}\lim_{N\rightarrow\infty
}\left(  \prod_{n=1}^{m-1}B(u(\frac{n}{N}\tau),du(\frac{n}{N}\tau))\right)
_{\nu}^{\mu}dy^{\nu}(m\epsilon)+
\end{align*}
which defines a global transformation from the rest system to the observer's
reference frame. Note that the $u(\frac{n}{N}\tau)$, $d(\frac{n}{N}\tau)$ and
$dy^{\nu}(n\epsilon)$ are four-velocities, their differential and the change
in the four coordinates of the particle, at the proper time value $\frac{n}%
{N}\tau,$ that is at an intermediate point of the trajectory. \ Thus, their
values define contributions to the total coordinate $y^{\mu}(\tau)$ of the
particle at the proper time $\tau,$ which "should" be transformed by all the
infinitesimal \ transformations \ ahead of the time $n\epsilon,$ to define
their contributions to the total coordinate of the particle $y^{\mu}(\tau).$
\ The product of the boosts is assumed to be ordered, with the index $n$
growing indicating the position from left to right and $u^{\mu}(\frac{N}%
{N}\tau)=u^{\mu}(\tau)=\left(
\begin{tabular}
[c]{l}%
1\\
$\overrightarrow{0}$%
\end{tabular}
\ \right)  .$ \ 

Now, the force in the observer's frame is given by the Lorentz transformation
of the force as defined in the rest frame. That is by the formula%
\begin{equation}
\mathcal{F}^{\mu}(\tau)=\Lambda_{\text{ \ }\nu}^{\mu}(\tau)\text{ }%
f_{e}^{\text{\ }\nu}(\tau),
\end{equation}
in which $\Lambda_{\nu}^{\mu}(\tau)$ is defined by%
\[
\Lambda_{\text{ \ }\nu}^{\mu}(\tau)=\lim_{N\rightarrow\infty}\left(
\prod_{n=1}^{N}B(u(\frac{n}{N}\tau),du(\frac{n}{N}\tau))\right)  _{\nu}^{\mu
}.
\]

This formula for the forces gives them a covariant definition. It can be
argued as follows. Let us consider the same construction of the transformation
$\Lambda_{\text{ \ }\nu}^{\mu}(\tau)$ \ but defined in another arbitrary
observer's frame, and denote it as $\widetilde{\Lambda}_{\text{ \ }\beta
}^{\alpha}(\tau).$ Then, the expressions of the forces in the two considered
frames are given as%

\begin{align*}
\mathcal{F}^{\mu}(\tau)  &  =\Lambda_{\text{ \ }\nu}^{\mu}(\tau)\text{ }%
f_{e}^{\text{\ }\nu}(\tau),\\
\widetilde{\mathcal{F}}^{\mu}(\tau)  &  =\widetilde{\Lambda}_{\text{ \ }\nu
}^{\mu}(\tau)\text{ }f_{e}^{\text{\ }\nu}(\tau),
\end{align*}
and after multiplying by the inverses of \ $\Lambda_{\text{ \ }\beta}^{\alpha
}(\tau)$ and $\widetilde{\Lambda}_{\text{ \ }\beta}^{\alpha}(\tau),$ if follows%

\begin{align*}
\Lambda_{\nu}^{\text{ \ }\mu}(\tau)\mathcal{F}^{\nu}(\tau)  &  =\text{ }%
f_{e}^{\text{\ }\mu}(\tau),\\
\widetilde{\Lambda}_{\nu}^{\text{ }\mu}(\tau)\widetilde{\mathcal{F}}^{\nu
}(\tau)  &  =\text{ }f_{e}^{\text{\ }\mu}(\tau),\\
\widetilde{\Lambda}_{\nu}^{\text{ }\mu}(\tau)\widetilde{\mathcal{F}}^{\nu
}(\tau)  &  =\Lambda_{\nu}^{\text{ \ }\mu}(\tau)\mathcal{F}^{\nu}(\tau),\\
\widetilde{\mathcal{F}}^{\mu}(\tau)  &  =\widetilde{\Lambda}_{\alpha}^{\text{
\ }\mu}(\tau)\Lambda_{\nu}^{\text{ \ }\alpha}(\tau)\mathcal{F}^{\nu}(\tau),\\
&  =\widehat{\Lambda}_{\nu}^{\text{ \ }\mu}(\tau)\mathcal{F}^{\nu}(\tau)
\end{align*}
which indicates that the defined forces in two arbitrary observer's frame are
related by the Lorentz transformation $\widehat{\Lambda}_{\nu}^{\text{ \ }\mu
}(\tau)$ linking both reference systems. Therefore, the force is defined as a
Lorentz vector and the \ following covariant Newton-like equation will be
considered
\begin{equation}
a^{\mu}(\tau)=\frac{1}{m}\mathcal{F}^{\mu}(\tau). \label{secondRel}%
\end{equation}

The connection of this equation with the ALD one will be discussed in the
following subsection. It should be noted, that when the motion is defined by a
force that does not maintain the velocities of the particles along a definite
direction, the analytic form of the force becomes complicated to determine.
This is due to the fact that the Lorentz boosts associated to velocities
oriented in different directions do not commute. This makes the analytic
determination more difficult.

For the explicit solution of the examples considered here, in which the motion
is collinear, the forces can be explicitly written, since the set of Lorentz
boosts along a fixed direction is an abelian group, which elements are given
in \ (\ref{boost}). Then, a Lorentz transformation $\Lambda_{\nu}^{\mu}$
\ expressing the coordinates of the observer's frame in terms of the rest one
in this simpler case can be chosen in the form
\begin{equation}
\Lambda_{\text{ \ }\nu}^{\mu}(\tau)\equiv\left(
\begin{tabular}
[c]{ll}%
$\gamma$ & $\gamma$ $v^{j}$\\
$\gamma$ $v^{i}$ & $\delta^{ij}+(\gamma-1)\frac{v^{i}\text{ }v^{j}}{v^{2}}$%
\end{tabular}
\ \ \right)  ,
\end{equation}
and correspondingly the formula for the force becomes
\begin{align}
\mathcal{F}^{\mu}(\tau)  &  =\Lambda_{\text{ \ }\nu}^{\mu}(\tau)\text{ }%
f_{e}^{\text{\ }\nu}(\tau)\nonumber\\
&  =(\gamma(v)\overrightarrow{v}(\tau).\overrightarrow{f_{e}}(\tau),\text{
}\overrightarrow{f_{e}}(\tau)+(\gamma-1)\frac{\overrightarrow{v}\text{
}\overrightarrow{f_{e}}(\tau)}{v^{2}}\overrightarrow{v}). \label{collinear}%
\end{align}
Below we simply enumerate some properties and conventions that can be helpful
to specify for what follows. The previous discussion determines that in the
rest frame of the particle, these relation are valid
\begin{equation}
a^{\mu}(\tau)u_{\mu}(\tau)=0,\text{ \ \ \ }a^{0}(\tau)=0.
\end{equation}

In this same rest system, the explicit form of the projection operator over
the three-space being orthogonal to the four-velocity is%
\begin{equation}
P^{\mu\nu}=g^{\mu\nu}-u^{\mu}u^{\nu}=\left\{
\begin{array}
[c]{cccc}%
0 & 0 & 0 & 0\\
0 & -1 & 0 & 0\\
0 & 0 & -1 & 0\\
0 & 0 & 0 & -1
\end{array}
\right\}  .
\end{equation}

For the sake of definiteness, let us explicitly write few basic relations
definning the spatial velocities and the connections between the proper and
observer's times
\begin{align}
\overrightarrow{v}(\tau) &  \equiv\frac{d}{dt}x^{i}(\tau),\text{
\ \ }i=1,2,3,\\
x^{0} &  =t,\\
d\tau &  =\sqrt{1-v^{2}}dt.
\end{align}

\subsection{ The satisfaction of the relativistic ALD equations}

Now, consider that the above defined Newton equations have a well defined
trajectory solving them, and then, study the question about: up to what point
this solution could also satisfies the ALD equations?. For this purpose, let
us evaluate the time derivative of the acceleration in the proper frame
%

\begin{equation}
\frac{d}{d\tau}a^{\mu}(\tau)=\frac{1}{m}\frac{d}{d\tau}\mathcal{F}^{\mu}%
(\tau)=\frac{1}{m}\frac{d}{d\tau}\Lambda_{\text{ \ }\nu}^{\mu}(\tau)\text{
}f_{e}^{\text{\ }\nu}(\tau),
\end{equation}
by considering the definition of $\Lambda_{\text{ \ }\nu}^{\mu}(\tau)$ as
follows%
\begin{align*}
\frac{d}{d\tau}(\Lambda_{\text{ }\nu}^{\mu}(\tau)f_{e}^{\text{\ }\nu}(\tau))
&  =\frac{d}{d\tau}\left(  \lim_{\epsilon\rightarrow\infty}\left(  \prod
_{n=1}^{\frac{\tau}{\epsilon}}B(u(n\text{ }\epsilon),du(n\text{ }%
\epsilon))\right)  _{\text{ \ }\nu}^{\mu}f_{e}^{\text{\ }\nu}(\tau)\right) \\
&  =\Lambda_{\text{ \ }\nu}^{\mu}(\tau)\lim_{\delta\rightarrow0}\left(
\frac{B(u(\tau+\delta),du(\tau+\delta))_{\text{ }\nu}^{\mu}f_{e}^{\text{\ }%
\nu}(\tau+\delta)-f_{e}^{\text{\ }\nu}(\tau)}{\delta}\right) \\
&  =\Lambda_{\text{ \ }\nu}^{\mu}(\tau)\lim_{\delta\rightarrow0}\left(
\frac{B(u(\tau+\delta),du(\tau+\delta))_{\text{ }\nu}^{\mu}f_{e}^{\text{\ }%
\nu}(\tau+\delta)-f_{e}^{\text{\ }\nu}(\tau)}{\delta}\right) \\
&  =\Lambda_{\text{ \ }\alpha}^{\mu}(\tau)\left(  {\large (}-u^{\alpha}%
(\tau)\frac{d}{d\tau}u(\tau)_{\nu}+\frac{d}{d\tau}u^{\alpha}(\tau)\text{
}u(\tau)_{\nu}{\large )}f_{e}^{\text{\ }\nu}(\tau)+\frac{d}{d\tau}%
f_{e}^{\text{\ }\alpha}(\tau)\right)  ,
\end{align*}
where all the quantities at the right of $\Lambda_{\text{ \ }\alpha}^{\mu
}(\tau)$ in the last line are defined in the rest frame. But, as mentioned
before, in the systems of coordinates
\[
u^{\mu}(\tau)=\left(
\begin{tabular}
[c]{l}%
$1$\\
$\overrightarrow{0}$%
\end{tabular}
\ \right)  .
\]

After recalling that also in this frame
\begin{equation}
\frac{d}{d\tau}(\gamma(v))|_{v\text{=}0}=-\frac{1}{2}\gamma^{\frac{3}{2}%
}(0)(2v\frac{d}{d\tau}v)|_{v\text{=}0}=0,
\end{equation}
it follows for the derivative of the acceleration
\[
\frac{d}{d\tau}u^{\mu}(\tau)=\left(
\begin{tabular}
[c]{l}%
$0$\\
$\frac{d}{d\tau}\overrightarrow{v}$%
\end{tabular}
\right)  .
\]

Using these relations it is possible to write
\begin{align*}
\frac{1}{m}\frac{d}{d\tau}(\Lambda_{\text{ \ }\nu}^{\mu}(\tau)f_{e}%
^{\text{\ }\nu}(\tau))  &  =\frac{\Lambda_{\text{ \ }\alpha}^{\mu}(\tau)}%
{m}\left(  -u^{\alpha}(\tau)\frac{d}{d\tau}u(\tau)_{\nu}f_{e}^{\text{\ }\nu
}(\tau)+\frac{d}{d\tau}f_{e}^{\text{\ }\alpha}(\tau)\right) \\
&  =\frac{\Lambda_{\text{ \ }\alpha}^{\mu}(\tau)}{m}\left(  u^{\alpha}%
(\tau)a^{i}(\tau)a^{i}(\tau)+\delta_{i}^{\alpha}\frac{m}{\kappa}(m\text{
}a^{i}(\tau)-f^{i}(\tau)\right) ,
\end{align*}
in which it had been employed the property of the considered force given in
(\ref{property}). After taking into account that $a^{i}(\tau)a^{i}%
(\tau)=-a^{\mu}(\tau)a_{\mu}(\tau)$, where $a^{i}(\tau)$ are the spatial
components of the acceleration in the rest frame and the temporal one
vanishes, \ it follows
\[
\frac{1}{m}\frac{d}{d\tau}(\Lambda_{\text{ }\nu}^{\mu}(\tau)f_{e}%
^{\text{\ }\nu}(\tau))=\Lambda_{\text{ }\alpha}^{\mu}(\tau){\large (-}%
u^{\alpha}(\tau)a^{\mu}(\tau)a_{\mu}(\tau)+\delta_{\text{ \ }i}^{\alpha}%
\frac{1}{\kappa}(m\text{ }a^{i}(\tau)-f^{i}(\tau)).
\]

But, since $u^{\alpha}(\tau)$\ is the four-velocity of the particle in the
rest frame, the vector \ $\Lambda_{\alpha}^{\mu}(\tau)u^{\alpha}(\tau)$\ \ is
its four-velocity in the observer's frame, and the previous expression can be
expressed in the form%

\[
\kappa(\frac{d}{d\tau}a^{\mu}(\tau)+u^{\mu}(\tau)a^{\nu}(\tau)a_{\nu}%
(\tau))=m\text{ }a^{\mu}(\tau)-f^{\mu}(\tau).
\]

Therefore, it follows that the satisfaction of the proposed Newton-like
equations implies the corresponding satisfaction of the Abraham-Lorentz-Dirac
ones, also in the relativistic case.\ We have directly checked the
satisfaction of the ALD equation for the case of \ the collinear motion in
which the force \ is explicitly defined by (\ref{collinear}). \ The derivation
of a formula for the general expression of the force is expected to be
considered elsewhere. \ The explicit solutions in the coming sections all
refer to the collinear motions. In addition, the proposal for a generalization
of the ALD equations for forces having sudden but finite changes, is also
restricted to this case, although its generalization seems to be possible.

\section{Forces defined in C$^{\infty}$}

Let us consider the series defining the components of the effective forces in
the rest frame system in the form
\begin{equation}
S_{\infty}^{\text{\ }i}(\tau)=\sum_{m=0}^{\infty}\frac{d^{m}}{d\tau^{m}}%
f^{i}(\tau)(\frac{\kappa}{m})^{m},
\end{equation}
and consider the three functions $f^{i}(t)$ as pertaining to the space of
infinitely smooth functions $C^{\infty}$ with the additional condition that
the series converges in an open region of \ proper times values. \ Let us
argue below that the set of all such series is not vanishing, and moreover
that it is a large class of functions. For this purpose, let us decompose the
series in a sum over a finite number of terms up to a largest index $m=$
$m_{f},$ plus the rest of the series%
\begin{equation}
S_{\infty}^{\text{\ }i}(\tau)=\sum_{m=0}^{m_{f}}\frac{d^{m}}{d\tau^{m}}%
f^{i}(\tau)(\frac{\kappa}{m})^{m}+\sum_{m=m_{f}}^{\infty}\frac{d^{m}}%
{d\tau^{m}}f^{i}(\tau)(\frac{\kappa}{m})^{m}.
\end{equation}
Then, let us assume that the only restriction on the functions $f^{i}(\tau)$
is that their times derivatives of arbitrary order, and at any time within the
mentioned open region, are bounded by a constant $M,$ for all the orders
higher that a given number $m(M).$ Then, let us select $m_{f}=m(M)$ which
allows to write the inequalities%
\begin{align}
|S_{\infty}^{\text{\ }i}(\tau)|  &  \leq|\sum_{m=0}^{m_{f}}\frac{d^{m}}%
{d\tau^{m}}f^{i}(\tau)(\frac{\kappa}{m})^{m}|\text{ }+\text{ }|\sum_{m=m_{f}%
}^{\infty}\frac{d^{m}}{d\tau^{m}}f^{i}(\tau)(\frac{\kappa}{m})^{m}|\nonumber\\
&  \leq|\sum_{m=0}^{m_{f}}\frac{d^{m}}{d\tau^{m}}f^{i}(\tau)(\frac{\kappa}%
{m})^{m}|\text{ }+M\text{ }|\sum_{m=m_{f}}^{\infty}(\frac{\kappa}{m}%
)^{m}|\nonumber\\
&  =|\sum_{m=0}^{m_{f}}\frac{d^{m}}{d\tau^{m}}f^{i}(\tau)(\frac{\kappa}%
{m})^{m}|\text{ }+M(\frac{\kappa}{m})^{m_{f}}|\frac{1}{1-\frac{\kappa}{m}%
}<\infty.
\end{align}

Thus, the series defining the effective forces are convergent at all time
values, with the unique condition that the time derivatives of arbitrary order
of the external forces are uniformly bounded for all orders. This constraint
seems not to be a strong one. By example, it is known, \ that when all the
times derivatives of a given function at a point are bounded, the functions
admits a Taylor expansion that converges to the value of the function in a
neighborhood of the considered point. That is, the class of external forces
for which the effective forces are well defined, includes a large set of
analytic functions.

\subsection{An analytic regularization of a constant force pulse}

In this section we will solve the effective Newton equations for an external
force \ which constitutes a regularization of a rigorously constant force
acting only during a specified time interval of duration $T$ and exactly
vanishing outside this time lapse. The degree of the regularization will be
defined by a time interval $t_{o}$ assumed to be very much shorter than $T$ .
\ It will be found that the parameter $t_{o}$ can be as shorter as ten times
the extremely short characteristic time $\frac{\varkappa}{m}$ being associated
to the radiation reaction forces in the ALD equations. However, in order to
numerically evidence the exact satisfaction of the ALD equations by the
solutions found for the effective Newton equations, larger values of
$\frac{\varkappa}{m}$ will be assumed. This will avoid the extremely small
values of the terms entering the series defining the effective forces, when
the "electromagnetic" values of $\frac{\varkappa}{m}$ are assumed. The
considered form of the force represents a regularization of the one employed
by Dirac in his classical work \begin{figure}[h]
\begin{center}
\hspace*{-0.4cm} \includegraphics[width=7.5cm]{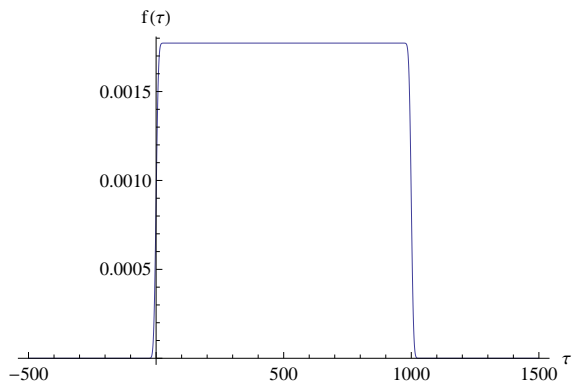}
\end{center}
\caption{ The plot illustrates the form of the external force defined by an
analytical function of the proper time along the whole real axis. Note that
the force can be seen as an infinitely differentiable regularization of an
exact square pulse showing the discontinuities at the times $\tau=0\,\,$s and
$\tau=1000\,\,$s .}%
\label{pulse1}%
\end{figure}\cite{Dirac}, in order to illustrate the appearance of runaway
solutions in the ALD equations. For the values of the parameters giving a
force in the form \ square pulse, it will be shown that the solution predicted
by the effective Newton equations does not show the runaway, nor the
pre-accelerated behavior, exhibited by the solutions derived by Dirac.
However, it will be also numerically checked that these solutions also satisfy
the ALD equations.

The explicit form of the "regularized" pulse defining the force in the rest
frame of the particle will be
\begin{align}
f(\tau,t_{o},T)  &  =f_{o}\int_{0}^{T}ds\exp(-\frac{(\tau-s)^{2}}{f_{1}%
^{2}t_{o}^{2}})\nonumber\\
&  =5\sqrt{\pi}t_{o}\text{ }f_{o}(Erf\text{ }(\frac{(T-\tau)}{f_{1}\text{
}t_{o}})+Erf\text{ }(\frac{\tau}{f_{1}\text{ }t_{o}})),\label{force}\\
f_{o}  &  =\frac{1}{10000},\text{ \ \ }f_{1}=10. \label{pulseF}%
\end{align}
where $Erf$ is the Error Function. \ The defined force is depicted in figure
\ref{pulse1} for the chosen values of \ $t_{o}=1$ cm and $T$ $=$1000 cm. note
that we are expressing the time in normal units. \ \ Let us no consider the
series defining the effective force $f_{e}(\tau),$ \ which is associated to
the external force $f(\tau):$%
\begin{equation}
f_{e}(\tau)=S_{\infty}^{\text{\ }}(\tau)=\sum_{m=0}^{\infty}\frac{d^{m}}%
{d\tau^{m}}f(\tau)\text{ }(\frac{\kappa}{m})^{m}.
\end{equation}

\begin{figure}[h]
\begin{center}
\hspace*{-0.4cm} \includegraphics[width=7.5cm]{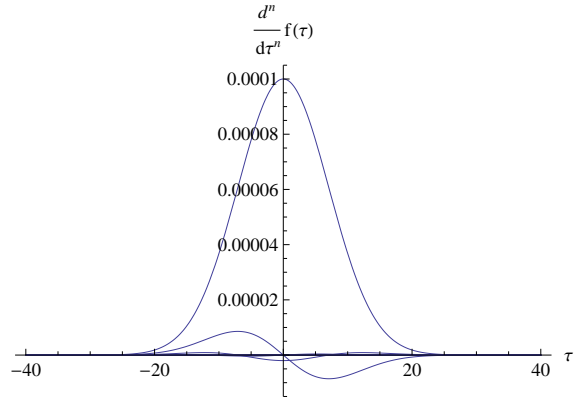}
\end{center}
\caption{ The figure shows the plots of the proper time derivatives up to the
order ten, of the force function associated to the defined analytic
regularization of the square pulse. The plot was done in a range of the origin
of coordinates in which the derivatives show the higher values, since the
pulse is transiting to attain its constant non vanishing value. As it can be
observed, all these derivatives are bounded functions of the proper time. Even
more, the bounds for the derivatives decrease with their order. This behavior
is maintained up to high values of the derivatives of order 50, where the
numerical errors started distorting the numerical results. }%
\label{derivft1}%
\end{figure}\bigskip

But, given $\frac{\kappa}{m}<1,$ which is extremely well satisfied by the case
of the electromagnetic ALD equation, the series defining the effective force
will converge just by only requiring that the time derivatives of arbitrary
order are uniformly bounded for all time values. The satisfaction of this
condition for the specific form to be considered for the force, after fixing
$t_{o}=1$ cm and $T$ $=1000$ cm, is evidenced in figure \ref{derivft1}. It
shows the plots of the times derivatives of orders $n=1,2,3,...,9,10.$ It can
be observed that all the depicted times derivatives are bounded, and moreover
the bound decreases when the order of the derivatives increases. This behavior
is maintained for higher orders, up to values in which the numerical precision
becomes degraded in our evaluation.

However, before assuming the form of the force in (\ref{pulseF}) by fixing the
values of $t_{o}$ and $T$ , \ it can be argued that the effective pulse like
forces are also well defined for extremely short "rising" times $t_{o}$ and
arbitrarily large time lapses $T$ of the pulses. This property can be argued
after performing the changes of variables
\begin{equation}
\tau=t_{o}\text{ }x,\text{ \ }s=t_{o}\text{ }y,
\end{equation}
which allows to express the force $f$ in the form
\begin{align}
f(\tau,t_{o},T)  &  =f^{\ast}(x,t_{o},\frac{T}{t_{o}})\nonumber\\
&  =f_{o}t_{o}\int_{0}^{\frac{T}{t_{o}}}dy\exp(-(\frac{x-y}{f_{1}}%
)^{2})\nonumber\\
&  =5\sqrt{\pi}t_{o}\text{ }f_{o}(Erf(\frac{(\frac{T}{t_{o}}-x)}{f_{1}\text{
}})+Erf(\frac{x}{f_{1}\text{ }})).
\end{align}

But, the implemented change of variables, allows to write for the effective
force series%
\begin{align}
f_{e}(\tau)  &  =\sum_{m=0}^{\infty}\frac{d^{m}}{dt^{m}}f(t,t_{o}%
,T)(\frac{\kappa}{m})^{m},\nonumber\\
&  =\sum_{m=0}^{\infty}\frac{d^{m}}{dx^{m}}f^{\ast}(x,t_{o},\frac{T}{t_{o}%
})(\frac{\kappa}{m}\frac{1}{t_{o}})^{m}. \label{f*}%
\end{align}

The last line of this relation, again indicates that with the "rising" time
being so short as to merely satisfying
\begin{equation}
t_{o}>\frac{\kappa}{m},
\end{equation}

\begin{figure}[h]
\begin{center}
\hspace*{-0.4cm} \includegraphics[width=7.5cm]{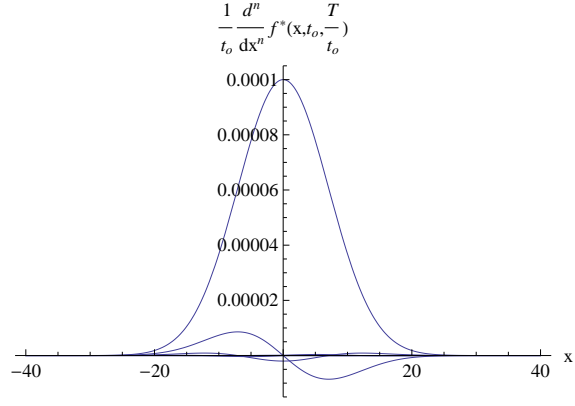}
\end{center}
\caption{ The behavior of the derivatives respect to the variable $x$ of the
auxiliary functions $f^{\ast}$ . The plots show derivatives up to the order
ten of the function $f^{\ast}$ being associated to the defined analytic
regularization of the square pulse. The graphic corresponds to a range of the
origin of coordinates in which the derivatives show the higher values. Again
all these derivatives are bounded functions of the coordinate $x$. }%
\label{derivfx2}%
\end{figure}\noindent the effective force as a function of the time (or
equivalently the variable $x$) becomes well defined if the arbitrary
derivatives over $x$ of the function $\ f^{\ast}(x,t_{o},\frac{T}{t_{o}})$ are
uniformly bounded for all the considered values of the variable $x$. But,
figure \ref{derivfx2} illustrates that the values of these derivatives as
functions of $\ x$ \ up to order ten, are bounded at all the values of the $x$
variable. This behavior is valid up to high orders for which the numerical
precision of the evaluations start to become degraded. Therefore, the results
indicate that the regularized constant force pulses, constructed in the
described analytic way, well define effective forces for very fast pulses,
which can rise so rapidly as few times the ALD time constant of nearly
10$^{-24}$ seconds in the electromagnetic case.

\subsection{Solutions of the Newton equations for the regularized pulsed
force}

The equations for the coordinates as functions of the proper time for the
pulsed force with the times parameters $t_{o}=1$ cm and $T$ $=1000$ cm will be
now numerically solved. With the use of their definition (\ref{secondRel})
these equations can be written in the form
\begin{align}
a^{\mu}(\tau)  &  =\frac{1}{m}\mathcal{F}^{\mu}\mathcal{(\tau)},\\
\mathcal{F}^{\mu}\mathcal{(\tau)}  &  =(\gamma\overrightarrow{v}%
.\overrightarrow{f_{e}},\text{ }\overrightarrow{f_{e}}+(\gamma-1)\frac
{\overrightarrow{v}.\overrightarrow{f_{e}}}{v^{2}}\overrightarrow{v}),\\
\overrightarrow{f_{e}}  &  =\sum_{m=0}^{\infty}\frac{d^{m}}{dt^{m}}%
f(\tau,t_{o},T)(\frac{\kappa}{m})^{m}.
\end{align}

In this 2D case we have the definitions
\begin{align}
u^{\mu}(\tau)  &  \equiv(\gamma,\gamma\overrightarrow{v}),\\
a^{\mu}(\tau)  &  \equiv(\frac{d}{d\tau}\gamma,\frac{d}{d\tau}(\gamma
\overrightarrow{v})).
\end{align}

The \ following two Newton equations can be explicitly written in the form
\begin{align}
\frac{d}{d\tau}u(\tau)  &  =\frac{1}{m}\gamma\text{ }f_{e}(\mathcal{\tau)},\\
\frac{d}{d\tau}u^{0}(\tau)  &  =\frac{1}{m}\gamma\text{ }f_{e}(\mathcal{\tau
)}v(\tau),\\
\gamma(\overrightarrow{v})  &  =\sqrt{1-v^{2}},\text{ \ \ }v^{2}%
=\overrightarrow{v}.\overrightarrow{v}.
\end{align}

But employing the definitions of the spatial velocity $\overrightarrow{v}$,
4-velocity $u^{\mu}$ and the effective forces $f_{e}$
\begin{align}
\overrightarrow{v}(\tau)  &  =\frac{d}{dt}\overrightarrow{x}=\frac{d}{\gamma
d\tau}\overrightarrow{x}(\tau),\\
u^{\mu}  &  =\frac{d}{d\tau}x^{\mu}(\tau),\\
\overrightarrow{f_{e}}  &  =\sum_{m=0}^{\infty}\frac{d^{m}}{d\tau^{m}}%
f(\tau,t_{o},T)(\frac{\kappa}{m})^{m},
\end{align}
the equations for the position and the time describing the trajectories
$(t(\tau),$ $x(\tau))$ which solve the Newton equations become%
\begin{align}
\frac{d^{2}}{d\tau^{2}}x(\tau)  &  =\frac{1}{m}\frac{d}{d\tau}t(\tau)\text{
}\sum_{m=0}^{\infty}\frac{d^{m}}{d\tau^{m}}f(\tau,t_{o},T)(\frac{\kappa}%
{m})^{m},\nonumber\\
\frac{d^{2}}{d\tau^{2}}t(\tau)  &  =\frac{1}{m}\text{ }\frac{d}{d\tau}%
x(\tau)\sum_{m=0}^{\infty}\frac{d^{m}}{d\tau^{m}}f(\tau,t_{o},T)\text{ }%
(\frac{\kappa}{m})^{m}. \label{newton2D}%
\end{align}

These equations are now solved for the particular values of the parameters%
\begin{align}
t_{o}  &  =1\text{ \ \ cm,}\\
T  &  =1000\text{ \ cm.}%
\end{align}

\begin{figure}[h]
\begin{center}
\hspace*{-0.4cm} \includegraphics[width=7.5cm]{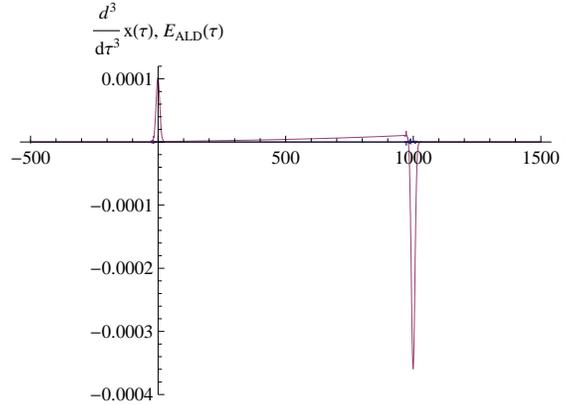}
\end{center}
\caption{The figure shows two plots in common: the value of the derivative of
the acceleration, which is the smaller term among the various contributions to
the ALD equations, and the function $E_{ALD}$ which vanishing implies the
satisfaction of the spatial ALD equation. The fact that the values of the
function $E_{ALD}$ can not be noticed in the plot, checks the very approximate
satisfaction of the spatial ALD equation }%
\label{ALDsatis5}%
\end{figure}

The constant $\frac{\kappa}{m}$ will be set to a value being in fact very much
higher than the one associated to the electron motion, which is nearly
10$^{-24}c$ cm. The chosen specific value $\frac{\kappa}{m}=0.8$ cm will help
to avoid extremely small higher orders contributions in powers of
$\frac{\kappa}{m}$, in the numerical solution of these equations. It can be
noticed that very much larger values of $\frac{\kappa}{m}$ with respect to the
one associated with the electron, are also of \ physical interest, by example
when considering the radiation of small moving objects in the air. The
solutions of the equations were considered for the following initial
conditions
\begin{align}
x(-50)  &  =0,\\
\frac{d}{d\tau}x(-50)  &  =0.2.
\end{align}
That is, at a proper time value of $-50$ cm , the particle is situated at the
origin of coordinates with a velocity given the proper time derivative of its
coordinates (the spatial component of the four-velocity) equal to $0.2$.
\ \ It can be noted that this problem is similar to the one considered by
Dirac to illustrate the appearance of pre-acceleration in the solutions of the
ALD equations \cite{Dirac}. The main difference between the two situations
\ is that here the pulse is not rigorously squared \ with discontinuous step
like transitions, but defined by an analytic function along the whole time
axis. \begin{figure}[h]
\begin{center}
\hspace*{-0.4cm} \includegraphics[width=11.5cm]{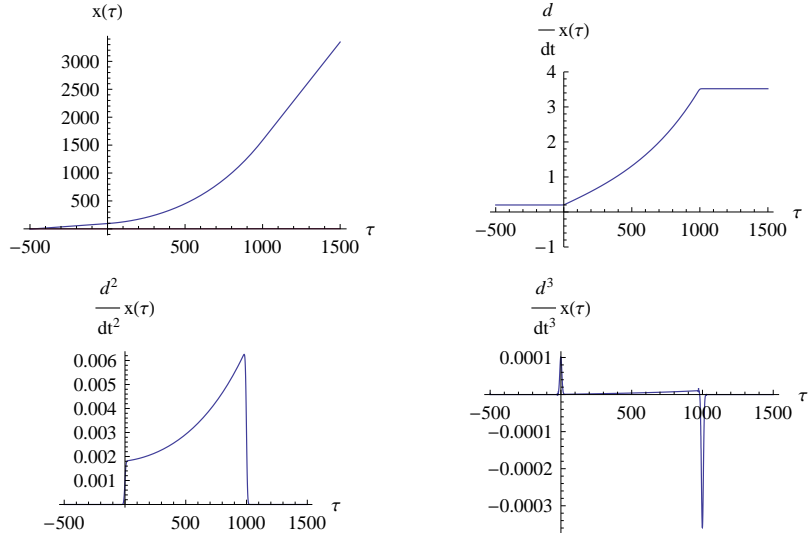}
\end{center}
\caption{The top left figure depicts the proper time evolution of the
coordinates of the particle upon which the force is acting. The top right one,
then shows the velocity of the particle in the same proper time interval. It
is clear that the motion tends to be uniform, outside of the interval
$(0,1000)$. The bottom left figure shows the behavior of the acceleration
which tends to vanishes outside the times in which the pulse gets appreciable
values. The last graph presents the dependence of the time derivative of the
acceleration. This quantity tends to be peaked around the instants at which
the pulse drastically changes its value. }%
\label{solution4}%
\end{figure}\bigskip The form of this pulse was shown in figure \ref{pulse1}.
The high value of the ratio $\frac{T}{t_{o}}$ gives to this function the
approximate pulse like appearance. The set of plots in figure \ref{solution4}
shows in first place, the proper time evolution of the coordinate of the
particle, indicating that the motion is nearly free before the time interval
of the pulse, for becoming accelerated for the times in which the force is
nearly constant. When the times is large and outside the region in which the
force is constant, the solutions becomes again a uniform motion as illustrated
by the vanishing of the acceleration in this zone. Note that the solution of
the Newton equations do not exhibit the pre-acceleration effect, nor the
runaway motions after the pulse is passed, as it was the case in the Dirac
solution of the ALD equations \cite{Dirac}. This is not a strange result given
that the effective force tends to vanish outside the pulse interval.
\ However, this example allows to numerically check that the obtained solution
also satisfies the ALD equations. This is clearly illustrated in figure
\ref{ALDsatis5}. It shows the plot of the following function, which vanishes
when the ALD equations are satisfied
\begin{equation}
E_{ALD}(\tau)=ma(\tau)-f(\tau)-\varkappa\text{ }(\frac{d}{d\tau}a(\tau
)+a^{\nu}(\tau)a_{\nu}(\tau)u(\tau)),
\end{equation}
in common with the plot of the time derivative of the acceleration
term of the ALD equations, which is the term of the equations showing, in
general, the smaller values along the times axis. As it can be noticed, the
values of the function $\ E_{ALD}(\tau)$ can not be noticed in comparison with
the values of the third derivative term. This indicates that the ALD equations
are satisfied within the precision of the numerical approximation of the
solution, confirming the general derivation presented before. Therefore an
interesting conclusion arises: the obtained solution of the effective Newton
equation, at the same time that also satisfies the ALD equation, \ avoids the
appearance of the pre-acceleration or runaways effects.

This example of an analytically regularized pulse, leads to an idea about how
justify a modification of the ALD equations for the case of non analytically
defined forces, presented in reference \cite{cabojorge}. This point will
discussed in next sections.

\section{\ Modified ALD equations for forces with sudden changes}

We will now assume that the existence of a sequence of forces $f_{k}^{\mu
}(\tau),$ $\ k=1,2,....\infty$ in all the proper time axis, which also show,
for each $k$, convergent values of the series defining their effective forces
$\ f_{e,k}^{\mu}(\tau),$ $\ k=1,2,....\infty$. \ \ A first general purpose of
this section will be to argue that, assumed that the sequence $f_{k}^{\mu
}(\tau)$ converges to a piecewise continuous limiting force $f^{\mu}(\tau)$,
the sequence of solutions of the corresponding effective Newton equations ALD
equations will tend to satisfy a set of modified ALD equations, which form
will be also determined. It will follow that these equations just generalize
the ones which were proposed in reference \cite{cabojorge}. \ In that work,
these equations were simply advanced under the basis of an idea: when a force
acting over a radiating classical electron is instantly removed, the Lienard
--Wiechert solution for the electromagnetic field surrounding the electron
within a sufficiently close neighborhood of its position, should be instantly
defined by a Lorentz boost transformed Coulomb field. But such fields, are
known to exert a vanishing four-force over the electron. This simple
observation strongly suggests that the mechanical equations driving the
electron motion, should be able to reproduce this effect: that is, to
instantaneously lead the acceleration of the particle to vanish, when a force
is removed in a extremely rapid way. \ The here found equations of motion
implement this property.

Let us firstly consider that the proper time axis is subdivided in a
denumerable set of contiguous \ intervals by the specific sequence of
increasing values of times
\begin{equation}
(\tau_{o},\tau_{o}^{(k)\ast},\tau_{1},\tau_{1}^{(k)\ast},\tau_{2},\tau
_{2}^{(k)\ast},...\tau_{n},\tau_{n}^{(k)\ast},.....).
\end{equation}

The general intervals ($\tau_{n},\tau_{n}^{(k)\ast})$ for arbitrary $n$ values
will be called as the "transition" intervals, in which the arbitrary time
derivatives for the forces of the sequence of $\ f_{k}^{\mu}(\tau)$ for all
$k$ values will be assumed as well defined, by also determining bounded values
for the effective forces $f_{e,k}^{\mu}(\tau)$ for all values of $k.$ \ In the
limit $k\rightarrow\infty$ we will assume that all the times $\tau_{n}$ not
showing the superindex $k$, \ will remain constant, and the other instant will
have limits $\tau_{n}^{(k)\ast}\rightarrow\tau_{n}.$ \ In the limit
$k\rightarrow\infty$ the sequence of forces will be assumed to possibly
approach a piecewise discontinuous limiting force at the points $\tau_{n}$ at
all $n$ values.

On the opposite way, within all the intervals $(\tau_{n}^{(k)\ast},\tau
_{n+1})$ the sequence of force functions will be assumed to tend to an
infinitely smooth function of the proper time, by also leading to well defined
effective forces. In particular if the force functions are given by polynomial
functions of the proper time with maximal order $N_{max}$, the series defining
the effective forces, having a finite number of terms, will always well define
effective forces.

Due to the assumption about the existence of the sequence of effective forces
along all the whole time axis for all finite $k$ values, the effective Newton
equations will be properly defined for all $k$ , and can be solved by simply
integrating over the proper time. We assume that the sequence of the effective
forces $f_{e,k}^{\mu}(\tau)$ will also tend to be piecewise continuous and
bounded at all the time axis. Then, this condition will imply that the time
integrals of the forces within all the transition intervals $(\tau_{n}%
,\tau_{n}^{(k)\ast})$ should vanish when these intervals shrink in the limit
$k\rightarrow\infty.$ \ \ Therefore, since integrals of this sort will define
the discontinuity in the four-velocities in the limit $k\rightarrow\infty,$ it
follows that the four-velocities should be continuous in the $k\rightarrow
\infty$ limit.

\subsection{ The modified ALD equations}

\ Consider now the sequence of the solutions of the of the effective Newton
equations $x^{(k)\mu}(\tau)$ for all values of $k$. From the previous
discussion, it is clear that for each $k$ value, the 4-velocities as functions
of the time will not tend to develop discontinuities across the transition
intervals when they reduce their sizes in the limit $k\rightarrow\infty.$
\ This follows, because the effective forces are assumed to exist and to be
also bounded. Thus, the impulses \ of these forces during the shrinking
\ transition intervals should tend to zero. Therefore, we will have that the
sequence of solutions tend to be piecewise smooth trajectory $x^{(k)\mu}%
(\tau)$ showing also a continuous velocity at the transition points in the
$k\rightarrow\infty$ limit.

Strictly inside all the intervals ($\tau_{n}^{(k)\ast},$ $\tau_{n+1})$ the
solutions $\ x^{(k)\mu}(\tau)$ can be constructed as satisfying the
$k\rightarrow\infty$ limit of the Newton equation, since the limiting
$k\rightarrow\infty$ values of the forces and the effective forces within
these intervals are assumed to be smooth and also bounded. Then, in these
zones they will also satisfy the ALD equation \ Therefore, within each of
these intervals the limiting $k\rightarrow\infty$ trajectory $x^{\mu}(\tau)$
should satisfy the effective Newton equations%

\begin{align}
ma^{\mu}(\tau)  &  =\mathcal{F}^{(n)\mu}\mathcal{(}\tau\mathcal{)}\nonumber\\
&  \equiv(\gamma(v)\overrightarrow{v}(\tau).\overrightarrow{f_{e}}^{(n)}%
(\tau),\overrightarrow{f_{e}}^{(n)}(\tau)+(\gamma-1)\frac{\overrightarrow{v}%
\text{ }\overrightarrow{f_{e}}^{(n)}(\tau)}{v^{2}}\overrightarrow{v}),
\label{efective}%
\end{align}
in which the effective forces $\ \ \overrightarrow{f_{e}}^{(n)}$\ are defined
by%
\begin{equation}
\overrightarrow{f_{e}}^{(n)}=\sum_{m=0}^{\infty}\frac{d^{m}}{dt^{m}%
}\overrightarrow{f}^{(n)}(\tau)(\frac{\kappa}{m})^{m},
\end{equation}
with $\overrightarrow{f}^{(n)}(\tau)$ being the limit $k\rightarrow\infty$ of
the sequence of forces $f_{k}^{\mu}(\tau)$ for the time taken within the
interval ($\tau_{n},\tau_{n+1}).$  At this point, as noted in the
introduction, we will restrict the discussion in this section to collinear
motions. The generalization to arbitrary motions seems to be feasible, and it
will be considered elsewhere.

However, due to the assumed developing of step like discontinuities of the
force values after the limit $k\rightarrow\infty,$ at the transition points
$\tau_{n},$ we can have for the effective four-force possibly different values
at the left $\tau_{n}^{-}$ and right $\tau_{n}^{+}$ of the transition
instants, with values%
\begin{align}
\mathcal{F}^{\mu}\mathcal{(}t_{n}^{-}\mathcal{)}  &  \equiv\left.
(\gamma(v)\overrightarrow{v}(\tau).\overrightarrow{f_{e}}^{(n-1)}%
(\tau),\overrightarrow{f_{e}}^{(n-1)}(\tau)+(\gamma-1)\frac{\overrightarrow{v}%
\text{ }\overrightarrow{f_{n}}^{(n-1)}(\tau)}{v^{2}}\overrightarrow{v}%
)\right\vert _{\tau=\tau_{n}^{-}},\label{cond1}\\
\mathcal{F}^{\mathcal{\mu}}\mathcal{(}t_{n}^{+}\mathcal{)}  &  \equiv\left.
(\gamma(v)\overrightarrow{v}(\tau).\overrightarrow{f_{e}}^{(n)}(\tau
),\overrightarrow{f_{e}}^{(n)}(\tau)+(\gamma-1)\frac{\overrightarrow{v}\text{
}\overrightarrow{f_{n}}^{(n)}(\tau)}{v^{2}}\overrightarrow{v})\right\vert
_{\tau=\tau_{n}^{+}}, \label{cond2}%
\end{align}
where, as noted before the coordinates and velocities are continuous at both
sides of the \ transition point $\tau_{n}.$ We will also define the force
functions inside each of the intervals $(t_{n},t_{n+1})$ as $\mathcal{F}%
^{(n)\mu}\mathcal{(}\tau\mathcal{)}=\mathcal{F}^{\mathcal{\mu}}\mathcal{(}%
\tau\mathcal{)}$ for $\tau$ $\epsilon$ $(\tau_{n},\tau_{n+1}).$ Thus
$\mathcal{F}^{\mu}\mathcal{(}\tau_{n}^{-}\mathcal{)}=\mathcal{F}^{(n)\mu
}\mathcal{(}\tau_{n}\mathcal{)}.$ Since these two limiting forces are equal to
the mass $m$ times de accelerations at any instant, \ this relation says that
the accelerations can be discontinuous at those instants. \ This property
defines a clear deviation from the case of the solutions showing
pre-acceleration or runaway behavior, which are assumed to have continuous
accelerations \ \cite{Dirac}. \ \ 

Then, let us now argue that the effective equations (\ref{efective}) \ with
the two boundary conditions (\ref{cond1}),(\ref{cond2}) are equivalent to the
set of modified ALD equations%
\begin{align}
ma^{\mu}(\tau)-f^{\mu}(\tau)  &  =\varkappa\text{ }(\frac{d}{d\tau}a^{\mu
}(\tau)+a^{\nu}(\tau)a_{\nu}(\tau)u^{\mu}(\tau))+\nonumber\\
&  +\frac{\varkappa}{m}\sum_{n=0}^{\infty}(\mathcal{F}^{\mu}\mathcal{(}%
t_{n}^{-}\mathcal{)}\delta^{(D,-)}(\tau-\tau_{n})-\mathcal{F}^{\mu}(\tau
_{n}^{+})\delta^{(D,+)}(\tau-\tau_{n}), \label{modec}%
\end{align}
where \ the Dirac Delta function, let say $\delta^{(D,-)}(\tau)$ \ \ is
defined as linear functionals determined by a sequence of functions of time
$\delta_{k}^{(D,-)}$($\tau),k=1,2,3,...,$ all having time integrals equal to
the unit for any $k$ value. Their supports for each $k$ is defined by
intervals $(-\epsilon_{k},0),$\ with positive $\epsilon_{k}$, which tend to
vanish in the limit $k\rightarrow\infty$. The linear functional acting over a
possibly piecewise continuous function $g(\tau)$ is then defined as the
integral%
\begin{equation}
\lim_{\tau\rightarrow0^{-}}(g(\tau))=\lim_{k\rightarrow\infty}\int d\tau\text{
}\delta_{k}^{(D,-)}(\tau)g(\tau).
\end{equation}

Therefore, if the function $g$ has a discontinuity of the step function type,
the integral of the function $\delta^{(D,-)}(\tau-\tau_{n})$ will give the
value of the limit of $g$ at the left of the point $\tau_{n}.$ The right Dirac
Delta $\delta^{(D,+)}(\tau)$ is defined in a similar way, but with support of
the form \ $(0$, $\epsilon_{k})$ with all $\epsilon_{k}$ again positive.

\bigskip

Now let us consider the acceleration $a^{\mu}(\tau)$ defined by the effective
force in a sufficiently small open neighborhood $B_{n}$ of each instant
$t_{n}.$ The expression for it can be written in the form%
\begin{equation}
a^{\mu}(\tau)=\frac{\mathcal{F}^{(n-1)\mu}\mathcal{(}\tau\mathcal{)}}{m}%
\Theta^{(-)}\mathcal{(}\tau_{n.}-\tau)\text{+}\frac{\mathcal{F}^{(n)\mu
}\mathcal{(}\tau\mathcal{)}}{m}\Theta^{(+)}\mathcal{(}\tau-\tau_{n.}),_{.}%
\end{equation}
where the special Heaviside like functions are defined as $\Theta^{(\pm
)}\mathcal{(}\tau)=\int_{-\infty}^{\tau}ds$ $\delta^{(D,\pm)}(s).$Therefore,
let us search for the equation satisfied by this expression for the
acceleration. \ The derivative of the acceleration within the neighborhood
$B_{n}$ takes the form
\begin{align}
\frac{d}{d\tau}a^{\mu}(\tau)  &  ={\Large (}\frac{d}{d\tau}(\frac
{\mathcal{F}^{(n-1)\mu}\mathcal{(}\tau\mathcal{)}}{m})\Theta^{(-)}%
\mathcal{(}\tau_{n.}-\tau)\text{+}\frac{d}{d\tau}(\frac{\mathcal{F}^{(n)\mu
}\mathcal{(}\tau\mathcal{)}}{m})\Theta^{(+)}(\tau-\tau_{n.})\nonumber\\
&  +\frac{\mathcal{F}^{(n-1)\mu}\mathcal{(}\tau\mathcal{)}}{m}\frac{d}{d\tau
}\Theta^{(-)}\mathcal{(}\tau_{n.}-\tau)\text{+}\frac{\mathcal{F}^{(n)\mu
}\mathcal{(}\tau\mathcal{)}}{m}\frac{d}{d\tau}\Theta^{(+)}(\tau-\tau
_{n.}){\Large )}\nonumber\\
&  ={\Large (}\frac{d}{d\tau}(\frac{\mathcal{F}^{(n-1)\mu}\mathcal{(}%
\tau\mathcal{)}}{m})\Theta^{(-)}\mathcal{(}\tau_{n.}-\tau)\text{+}\frac
{d}{d\tau}\frac{\mathcal{F}^{(n)\mu}\mathcal{(}\tau\mathcal{)}}{m}\Theta
^{(+)}(\tau-\tau_{n.})+\nonumber\\
&  -\frac{\mathcal{F}^{(n-1)\mu}\mathcal{(}\tau\mathcal{)}}{m}\delta
^{(D,-)}(\tau-\tau_{n})\text{+}\frac{\mathcal{F}^{(n)\mu}\mathcal{(}%
\tau\mathcal{)}}{m}\delta^{(D,+)}(\tau-\tau_{n}){\Large )}.
\end{align}

Henceforth, after substituting this expression in (\ref{modec}), and
considering that the ALD equations are satisfied at the interior points of all
the intervals $(\tau_{n},\tau_{n+1}),$ it follows%
\begin{align}
&  {\LARGE (}{\large (}ma^{\mu}(\tau)-f^{\mu(n-1)}(\tau)-\varkappa\text{
}(\frac{d}{d\tau}a^{\mu}(\tau)+a^{\nu}(\tau)a_{\nu}(\tau)u^{\mu}%
(\tau)){\large )}\mathcal{\Theta}^{(-)}\mathcal{(}\tau_{n.}-\tau)+\nonumber\\
&  {\large (}ma^{\mu}(\tau)-f^{\mu(n)}(\tau)-\varkappa\text{ }(\frac{d}{d\tau
}a^{\mu}(\tau)+a^{\nu}(\tau)a_{\nu}(\tau)u^{\mu}(\tau)){\large )}%
\mathcal{\Theta}^{(+)}\mathcal{(}\tau-\tau_{n.})\nonumber\\
&  -\frac{\varkappa\mathcal{F}^{(n-1)\mu}\mathcal{(}\tau\mathcal{)}}{m}%
\delta^{(D,-)}(\tau-\tau_{n})+\frac{\varkappa\mathcal{F}^{(n)\mu}%
\mathcal{(}\tau\mathcal{)}}{m}\delta^{(D,+)}(\tau-\tau_{n}).\nonumber\\
&  +\frac{\varkappa}{m}(\mathcal{F}^{\mu}\mathcal{(}\tau_{n}^{-}%
\mathcal{)}\delta^{(D,-)}(\tau-\tau_{n})-\mathcal{F}^{\mu}(\tau_{n}^{+}%
)\delta^{(D,+)}(\tau-\tau_{n}){\LARGE )}=0.
\end{align}

Thus, all the terms in the above equations add to zero around each transition
time $t_{n}$. The first two lines line vanish because the trajectories solving
the Newton equations within each of the intervals $(\tau_{n},\tau_{n+1})$ also
satisfy the ALD equations within each neighborhood $B_{n}$except at the point
$t_{n}$. The last two terms also cancels between themselves since the Delta
functions allow to evaluate the argument of the functions multiplying them at
their support point. Therefore, after considering that the modifying terms of
the ALD equations vanish outside all the vicinities $B_{n},$ \ \ it follows
that the limiting solution\ along the whole axis satisfies the modified ALD
equations
\begin{align}
ma^{\mu}(\tau)-f^{\mu}(\tau)  &  =\varkappa\text{ }(\frac{d}{d\tau}a^{\mu
}(\tau)+a^{\nu}(\tau)a_{\nu}(\tau)u^{\mu}(\tau))+\nonumber\\
&  +\varkappa\sum_{n=0}^{\infty}(a^{\mu}(\tau_{n}^{-})\delta^{(D,-)}(\tau
-\tau_{n})-a^{\mu}(\tau_{n}^{+})\delta^{(D,+)}(\tau-\tau_{n}).
\end{align}

\subsection{The solution for the constant force pulse}

\ Let us consider now the solution of the modified ALD equations for the case
of a exact squared pulse of the similar form as the before considered analytic
one. The same initial conditions for the position and velocities will be fixed%
\begin{align}
x(-50)  &  =0,\\
\frac{d}{d\tau}x(-50)  &  =0.2.
\end{align}

\begin{figure}[h]
\begin{center}
\hspace*{-0.4cm} \includegraphics[width=7.5cm]{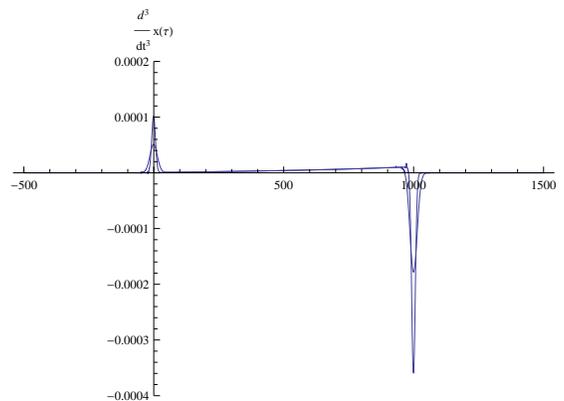}
\end{center}
\caption{ The superposed plot corresponds to two analytic pulsed forces of
nearly equal height and width, but one having a rising time double than the
other. One curve is related with the higher left and right peaks, and the
other is associated with lower ones at both sides. The curve showing the
higher peaks corresponds to the pulse with shorter rising time, which
illustrates how the Dirac' Delta functions appearing in the modified ALD
equations are gradually generated as the pulse approaches the exactly squared
form. }%
\label{spikes}%
\end{figure}

The Newton equations in the interior points of any of the three intervals in
which the time axis is decomposed by the two instants $\tau_{o}=0$ and
$\tau_{1}=T,$ have basically the same form as the equation (\ref{newton2D})
\begin{align}
\frac{ds^{2}}{d\tau^{2}}x(\tau)  &  =\frac{1}{m}\frac{d}{d\tau}t(\tau)\text{
}f_{P}(\tau,T),\nonumber\\
\frac{d^{2}}{d\tau^{2}}t(\tau)  &  =\frac{1}{m}\text{ }\frac{d}{d\tau}%
x(\tau)f_{P}(\tau,T), \label{newton2Dpulse}%
\end{align}
where the force now defining the exact pulse of constant amplitude is given by
the formula
\begin{equation}
f_{P}(\tau,T)=f^{p\text{ }}\Theta(\tau)\text{ }\Theta(T-\tau),
\end{equation}
in which $\Theta(\tau)$ is the Heaviside function. The width of the pulse was
chosen as given by the same parameter $T$ defining the width of the
analytically regularized pulse in past sections. The constant amplitude of the
pulse $f^{p\text{ }}$ will be approximately coinciding with the \ height of
the \ regularized pulse, by selecting its magnitude as
\begin{equation}
f^{p\text{ }}=f(\frac{T}{2},t_{o},T).
\end{equation}
That is, by the height of the pulse at a time equal to\ half of its
approximate width $T$ . The parameters for the numerical evaluation will also
coincide with values selected before \ $t_{o}=1$ and $\ T=1000.$
\begin{figure}[h]
\begin{center}
\hspace*{-0.4cm} \includegraphics[width=11.5cm]{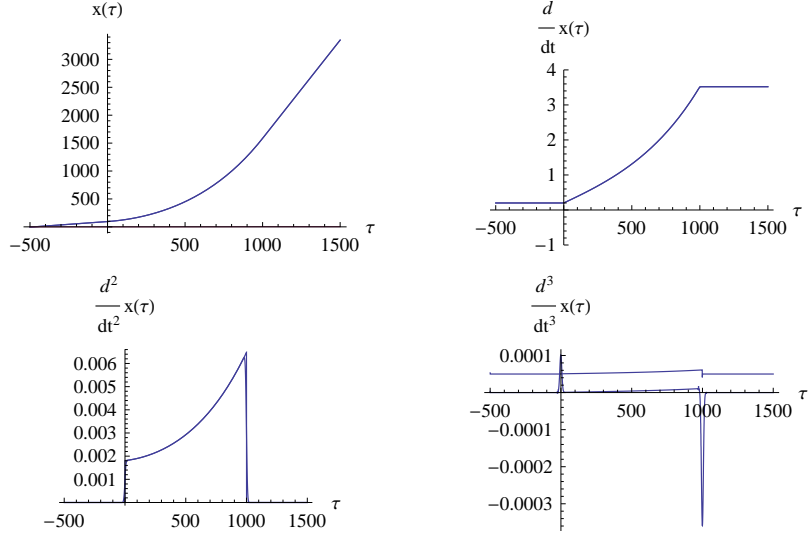}
\end{center}
\caption{ The array of figures show in each graphic two plots, one associated
to the analytic pulse of force and the other relate with the exact square
pulse solving the modified ALD equations. The top left figure shows the
coordinates of both solutions. Note the close coincidence of the coordinates.
The top right curves for the velocities also illustrates the similarity
between the solutions for this quantity. The bottom left plot present the
results for the accelerations for the two solutions. The non constant behavior
of the acceleration shows that the motion is relativistic. Finally the bottom
right plot compares the time derivatives of the acceleration for both
solutions. For better evidencing the difference, the quantity associated to
the exactly square pulse is shifted in a positive constant along the vertical
axis. It can be noted that in the internals points of the interval $(0,1000)$
the two evaluations tend to coincide. However, in the points close to the
boundaries of this interval, the derivative of the acceleration associated to
the analytic pulse develops peaked values. These peaked values gives account
of the developing of the Dirac functions entering the modified ALD equations.
The analytic solution, as obeying the exact ALD along the whole axis should
generate these Dirac functions, in order to imply the modified equations in
the limit.}%
\label{comparison}%
\end{figure}\bigskip The solution of the equations (\ref{newton2Dpulse}) for
the time dependence of the coordinate, velocity, acceleration and its time
derivative, \ are jointly plotted in figure \ref{comparison}, for both pulses:
the one of exactly constant amplitude and the analytically regularized one.
\ The plots evidence that these quantities are closely similar for both
forces. A small difference starts to be noticed in the curve for the
velocities. It shows that only in the close neighborhood of the transition
points, in which the forces suddenly change, the rapid but smooth transition
of the velocity for the \ regularized pulse can be observed. Further, the plot
of the derivative of the acceleration is also presenting a difference. \ For
making the comparison clearer, the plot associated to the derivative of the
acceleration for the pulse of exact constant amplitude, is slightly shifted
along the vertical axis away the zero values. \ This permits to note that the
magnitude of the derivative of the acceleration at the interior points of the
interval $(0,T)$ closely coincide for both solutions. However, near the
transition points, the regularized pulse presents a peaked behavior. \ This is
a numerical confirmation for the validity of the modified equations, since
these peaked dependence are necessary for reproducing the Dirac's delta
functions entering these modified equations, in the limit when the regularized
pulse tends to becomes the exactly constant one. That is, when the regularized
pulse is gradually made to be even more similar to the constant pulse, it can
be expected that the "spikes" appearing at the transition points for the
derivatives of the acceleration, are associated to regularizations of the
Dirac's distributions. The enhancing of the peaks when the regularized pulse
tends to be closer to the exact constant ones is illustrated in figure
\ref{spikes}. It shows the derivative of the acceleration for two solutions
with almost all the parameters identical to the ones considered before, but
differing only in the value determining the pulse rise time $t_{o}$.
\bigskip The values selected for this parameters were $t_{o}=1$ cm
and $t_{o}=2$ cm. \ That is, the pulse with $t_{o}=1$ will have a rising time
two times smaller than the one with $t_{o}=2$ m. In the picture, there is a
curve which shows the smaller peaks at the right and left of the figure, and
the other one presents the higher peaks at both sides. \ The curve with the
larger values is associated to the pulse $t_{o}=1$ and the curve with smaller
values is related with the slower rising pulse. Thus, it is clear that when
the analytic pulse tends to approach the exactly constant one, the solution
tends to show time derivatives with the appearance of \ regularizations of the
Dirac's Delta functions, as the it should be, if the modified ALD equations
are implied by the limit of the exact ALD equations.

The absence of the peaks in the solution linked with the square pulse is
associate with the fact that the equations were solved at the interior points,
and the boundary conditions are automatically satisfied across the two
transition points, by the solutions of the Newton equations in the three
intervals. These boundary conditions were argued before to be equivalent to
the presence of Dirac Delta functions with supports in the transition points.

\section*{Summary}

The work presents a second order Newton like equations of motion for a
radiating particle. It was argued that the trajectories obeying the equation
exactly satisfy the Abraham-Lorentz-Dirac (ALD) equations. Forces which only
depend on the proper time were considered by now. A condition for these
properties to occurs in a given time interval is derived: it is sufficient
that the force becomes infinitely smooth and also that a particular series
defined by the infinite sequence of its time derivatives converges to a
bounded function. This series defines in a local way the effective force
determining the Newton effective equation. The existing solutions of such
effective equations can not show the runaways or pre-acceleration phenomena.
The Newton equations were numerically solved for a pulsed like force given by
an analytic function on the whole proper time axis. The satisfaction of the
ALD equations by the obtained solution is numerically checked. In addition,
for the case of the collinear motions, it was derived a set of modified ALD
equations for almost infinitely smooth forces, which however, show step like
discontinuities. The form of these equations supported the statement argued in
a former work, about that the Lienard-Wiechert field surrounding a radiating
particle should determine that the effective force on the particle
instantaneously vanishes, when the external force is suddenly removed. The
modified ALD equations argued in the former study are here derived in a more
general form, in which a suddenly applied external force is also instantly
creating an effective non vanishing acceleration. The work is expected to be
extended in some directions. By example, one issue which seems of interest to
define is whether the class of external forces which also show well defined
effective forces, constitutes a dense subset (within an appropriate norm) at
least within the set of forces defined by continuous proper time functions.
This property could help to understand, whether the ALD equations, for any
continuous time dependent force, could always exhibit a solution, which is
also approximately solving second order Newton equations, and then not showing
pre-accelerated or runaway behavior.

\section*{Acknowledgments}

The authors would like to deeply acknowledge the helpful comments on the
subject of this work received from Danilo Villarroel (Chile), Jorge
Casti\~{n}eiras (UFPA, Brazil), Luis Carlos Bassalo Crispino (UFPA, Brazil)
and Suvrat \ Raju \ (ICTS, Tata Institute, India). The support also received
by both authors from the Caribbean Network on Quantum Mechanics, Particles and
Fields (Net-35) of the ICTP Office of External Activities (OEA) and the
\ "Proyecto Nacional de Ciencias B\'{a}sicas"(PNCB) of CITMA, Cuba is also
very much acknowledged.

\end{document}